\documentclass[manuscript]{aastex}

\slugcomment{To be submitted}

\shorttitle{The Farley-Buneman Instability in the Solar Chromosphere}
\shortauthors{Gogoberidze et al.}

\begin{document}


\title{Farley-Buneman Instability in the Solar Chromosphere}


\author{G. Gogoberidze\altaffilmark{1}, Y. Voitenko, S. Poedts\altaffilmark{2}, and M. Goossens}
\affil{CPA/K.U.Leuven, Celestijnenlaan 200B, 3001 Leuven, Belgium}
\altaffiltext{1}{Georgian National Astrophysical Observatory, Kazbegi ave. $2^a$ Tbilisi-0160, Georgia}
\altaffiltext{2}{Leuven Mathematical Modeling \& Computational Science Centre, Belgium}

\begin{abstract}
The Farley-Buneman instability is studied in the partially ionized
plasma of the solar chromosphere taking into account the finite
magnetization of the ions and Coulomb collisions. We obtain the
threshold value for the relative velocity between ions and
electrons necessary for the instability to develop. It is
shown that Coulomb collisions play a destabilizing role in the sense that
they enable the instability even in the regions where the ion
magnetization is greater than unity. By applying these
results to chromospheric conditions, we show that the Farley-Buneman
instability can not be responsible for the quasi-steady
heating of the solar chromosphere. However, in the presence of
strong cross-field currents it can produce small-scale, $\sim
0.1-3$ m, density irregularities in the solar chromosphere. These
irregularities can cause scintillations of radio waves with
similar wave lengths and provide a tool for remote
chromospheric sensing.
\end{abstract}





\section{Introduction}

The mechanism of chromospheric heating is a major puzzle of solar
physics since it was discovered that the temperature in the solar chromosphere is
much higher than what can be expected for a plasma in radiative equilibrium. The first
possible scenario for explaining of chromospheric heating was proposed by \citet{B46} and
\citet{S48}, who suggested that the inner atmosphere of the sun is heated by
acoustic waves that are generated in the convective zone. Later, theoretical and
numerical studies \citep{S67,CS92} have demonstrated that acoustic
waves are, in fact, abundantly generated in the convective zone and that these
waves can, in principle, be responsible for chromospheric heating. However,
measurements of the acoustic flux at different chromospheric levels usually fail
to find sufficient energy to heat the whole chromosphere \citep{FC05}.
However, as it has been shown recently, the chromosphere in the magnetic network may be heated
by magnetoacoustic waves generated locally, inside or in the vicinity of
the magnetic flux tubes \citep{HB08}.
Also random Alfven waves can heat upper chromosphere via ion-neutral collisions and
generate slow shocks, which can explain the formation of spicules \citep{EJ04}.

As an alternative explanation for the chromospheric heating,
it has been suggested \citep{P88,S99} that impulsive nano-flares,
powered by magnetic reconnection events, could be responsible
for chromospheric heating. Although the observations show numerous
transient brightenings on the sun, these are insufficiently
frequent and insufficiently energetic to explain the persistent UV emission of the chromosphere
\citep{A00}. During solar flares, the chromosphere can be strongly heated and ionized locally
by precipitating electron beams and evaporate upward, producing observed polarised $H\alpha$ emission
via collisional interaction with neutral surrounding hydrogen \citep{FB98}.

Yet another possibility for chromospheric heating is the resistive dissipation of electric currents
\citep{RM84,G04}. Recent analysis of three dimensional vector currents and
temperatures, deduced from spectropolarimetric observations of a
sunspot from photospheric to chromospheric levels, has shown that,
while resistive current dissipation can contribute to heat the
sunspot chromosphere, it is not the dominant factor \citep{S07}. Recently, it
has been suggested that the Farley-Buneman \citep{F63,B63}
instability (FBI), driven by convective motions, can be responsible for
chromospheric heating \citep{L00,F05,FPH08}. The FBI is known to create plasma irregularities in the
terrestrial ionospheric E-region, at heights where the electrons are
strongly magnetized. The interplay of the earth's electric and
geomagnetic field produces currents which give rise to the FB
instability. Similarly, in those places where the electrons are
strongly magnetized, the collisional drag of the ions by neutral
flows can cause the development of a similar instability. Using the decrement of the
FBI derived by \citep{F63} and assuming a negligible ion magnetization,
\citet{FPH08} concluded that the FBI should be present at least in the upper
half of the chromosphere. Earlier, the analysis of \citet{L00} had
indicated that the FBI might operate in the chromosphere at heights $h>1000~\mathrm{km}$.

However, the studies of the FBI in the chromosphere conditions are
incomplete and they do not take into account two effects which under
chromospheric conditions are important as we will show below.
Firstly, if the finite
magnetization of ions is taken into account, the Hall current perturbations
weaken the FBI, and the system becomes stable for any neutral
flow velocity when the ion magnetization factor $\kappa$
exceeds unity \citep{FPF84}. Therefore, this instability can not
operate in the upper solar chromosphere, where $\kappa>1$. Secondly, contrary
to the E-layer plasma in the Earth's atmosphere, and top of the solar photosphere \citep{PVP07},
the ionization degree in the solar
chromosphere is quite high ($10^{-2}-10^{-4}$) and, consequently, Coulomb collisions can not
be ignored as is usually done in the study of the E-layer plasma.

In this letter, we study the FBI taking into account both the finite
magnetization of ions and the Coulomb collisions. We shall show that,
in contrast to the situation in very weakly ionized plasmas,
the relatively high degree of ionisation in the solar
chromosphere makes the Coulomb collisions important for the FBI
development. As a result, the instability becomes possible even in plasmas with
an ion magnetization $\kappa$ larger than unity. However, by applying our
analytical results to the solar chromosphere, we show that even though the FBI can
sporadically appear in the chromosphere, it cannot be the main source of
chromospheric heating.

\section{Formalism}

We consider a weakly ionized plasma consisting of electrons, one species of singly charged
ions and neutral hydrogen.
In the upper solar chromosphere, the positively charged particles are mainly protons, whereas at
lower altitudes the positive charge is dominated by heavy ions. We
therefore do not further specify the type of ions, so that our results are applicable to
both the upper and the lower chromosphere.

The dynamics of electrons and ions in such plasmas is governed by the continuity and the Euler equations, viz.
\begin{equation}
\frac{\partial n_\alpha}{\partial t}+ \nabla \cdot (n_\alpha {\bf V}_\alpha)=0,
 \label{eq:1}
\end{equation}
and
\begin{equation}
m_\alpha\frac{{\rm d}_\alpha {\bf V}_\alpha}{\partial t} =q_\alpha \left( {\bf E}+\frac{{\bf V}_\alpha\times {\bf B}}{c} \right) -\frac{\nabla n_\alpha \mathcal{K} T_\alpha}{n_\alpha}- m_e \nu_{ep}({\bf V}_\alpha-{\bf V}_{\alpha^\ast})- m_\alpha \nu_{\alpha n}({\bf V}_\alpha-{\bf V}_n).
 \label{eq:2}
\end{equation}
Here, $\alpha=e,i$ denotes electrons or ions, $\alpha^\ast$ denotes the charged species opposite to $\alpha$, $\bar \alpha$ stands for $e$ for the electron equation and for proton (p) for the ion equation. Also, $n$ corresponds to neutrals, and $n_\alpha$ denotes the density, ${ \bf V}_\alpha$ is the averaged drift velocity, $m_\alpha$ is the mass, $T_\alpha$ is the temperature, $q_\alpha$ is the charge, $\nu_{\alpha \beta}$ is the collision frequency, $c$ is the speed of light, $\mathcal{K}$ is The Boltzmann constant, and ${\rm d_\alpha}/{\rm d}t $ denotes the convective derivative.

For the electron-ion and electron-neutron collision frequencies we use the following expressions \citep{B65}
\begin{equation}
\nu_{ei}=\frac{4(2\pi)^{1/2}e^4 n_e \Lambda }{3m_e^{1/2}(\mathcal{K}T)^{3/2}}, \label{eq:c1}
\end{equation}
and
\begin{eqnarray}
\nu_{en}=\sigma_{en}n_n \sqrt{ \frac{\mathcal{K}T_e}{m_e}},~\nu_{in}=\nu_{pn}=\sigma_{in}n_n \sqrt{\frac{\mathcal{K}T_p}{m_p} },\label{eq:c2}
\end{eqnarray}
where $\Lambda$ corresponds to the Coulomb logarithm and in the former equation we take into account
the fact that, regardless the mass of the dominant ion species, the ion-neutral
collision frequency in the solar chromosphere is proportional to the thermal velocity of
the neutral (hydrogen) component, because in the case of equal temperatures
the thermal velocity of the neutral particles equals to or is greater than
the thermal velocity of the ions, depending on the mass of the ions.
The electron-neutral and ion-neutral collision cross sections
are $\sigma_{en}=3.0\times10^{-15}~{\rm cm^2}$ \citep{BK71} and
$\sigma_{in}=2.8\times10^{-14}~{\rm cm^2}$ \citep{KS99}, respectively.

We assume, that the system is penetrated by a uniform magnetic field ${\bf B} $
and that the neutrals have a background velocity ${\bf V}_n \perp {\bf B}$. The Eqs.~(\ref{eq:2})
then yield a stationary solution for the background ion drift velocity
\begin{equation}
{\bf V}_i=\frac{\left( 1+ \kappa \kappa_1 \psi N  \right) {\bf V}_n + \kappa {\bf V}_n\times \hat z }{1+\kappa_1^2 } ,
 \label{eq:3a}
\end{equation}
and for the so-called `current velocity', i.e.\ the relative velocity between the ions and the electrons,
\begin{equation}
{\bf U}_0={\bf V}_i-{\bf V}_e=\frac{{\bf V}_n + \kappa_1  {\bf V}_n\times \hat z}{1+\kappa_1^2}.
 \label{eq:3}
\end{equation}
Here $\kappa=\omega_{cp}/\nu_{pn} $ is proton magnetization, $\omega_{c\alpha} \equiv eB/m_\alpha c$ is the cyclotron frequency, $\kappa_1=\kappa(1+\psi N) $,~ $\psi=\nu_{en} \nu_{in}/\omega_{cp}\omega_{ce}$ and $N=\nu_{ep}/\nu_{en}$ is the ratio of Coulomb and electron-neutral collision frequencies.

On this background, we study linear electrostatic perturbations in the plane
perpendicular to the background magnetic field. In order to simplify analysis we admit standard assumptions for
the study of the FBI in
the Earth's E-layer plasma \citep{OMO96,SN}. We assume quasi-neutrality ($n_e\approx n_i$).
Technically, this means that, instead of using Poisson's equation,
we use $\nabla {\bf J}=0$, where ${\bf J}$ denotes the electric current density.
In addition, because $n_e$ and $n_i$ are indistinguishable, we use only one continuity equation. Finally,
we treat the electrons as massless because the FBI occurs on an ion-neutral collision timescale
which, for typical chromospheric conditions, strongly exceeds both the electron cyclotron gyration and
the electron plasma oscillation timescales.

We Linearize Eqs.~(\ref{eq:1})-(\ref{eq:2}) and perform a Fourier transform of the obtained equations, and after long but straightforward algebra arrive at the following dispersion equation
\begin{equation}
\frac{\omega-  {\bf k} \cdot {\bf U}_0}{\psi}  +\frac{(1-i \omega/\nu_{in}^\prime)^2 +\kappa^2}{1- i \omega/\nu_{in}^\prime} \omega+N(1-i \omega/\nu_{in}^\prime)\omega+i(1+N)\frac{c_s^2k^2}{\nu_{in}^\prime}=0,
 \label{eq:4}
\end{equation}
where $c_s \equiv [\mathcal{K}(T_e+T_i)/m_i]^{1/2}$ is the sound velocity and $\nu_{in}^\prime=m_p \nu_{in}/m_i$. In the derivation of the dispersion equation, we neglected all terms of the order
of the small parameter $\psi \kappa^2 \sim m_e \nu_{en} /m_p \nu_{pn} \sim 2.6\times 10^{-3}$.

In the limit of low-frequency and long-wavelength perturbations ($
\left\vert \omega \right\vert ,~\left\vert \mathbf{k}\cdot \mathbf{U}%
_{0}\right\vert \ll \nu _{in}^{\prime }$), we obtain the oscillation frequency $\omega_r$
and the growth rate $\gamma$ of the Farley-Buneman type instability:
\begin{equation}
\omega_r= \frac{{\bf k} \cdot {\bf U}_0}{1+\bar \psi},
 \label{eq:5}
\end{equation}
and
\begin{equation}
\gamma = \frac{\bar \psi ({\bf k} \cdot {\bf U}_0)^2}{\nu_{in}^\prime (1+\bar \psi)}\left[ \frac{1-\kappa^2/(1+N)}{(1+\bar \psi)^2} -\frac{k^2 c_s^2}{({\bf k} \cdot U_0)^2} \right],
 \label{eq:6}
\end{equation}
where $\bar \psi \equiv \psi(1+N)$.

\section{Discussion and conclusions}

\begin{figure}
\epsscale{.80}
\plotone{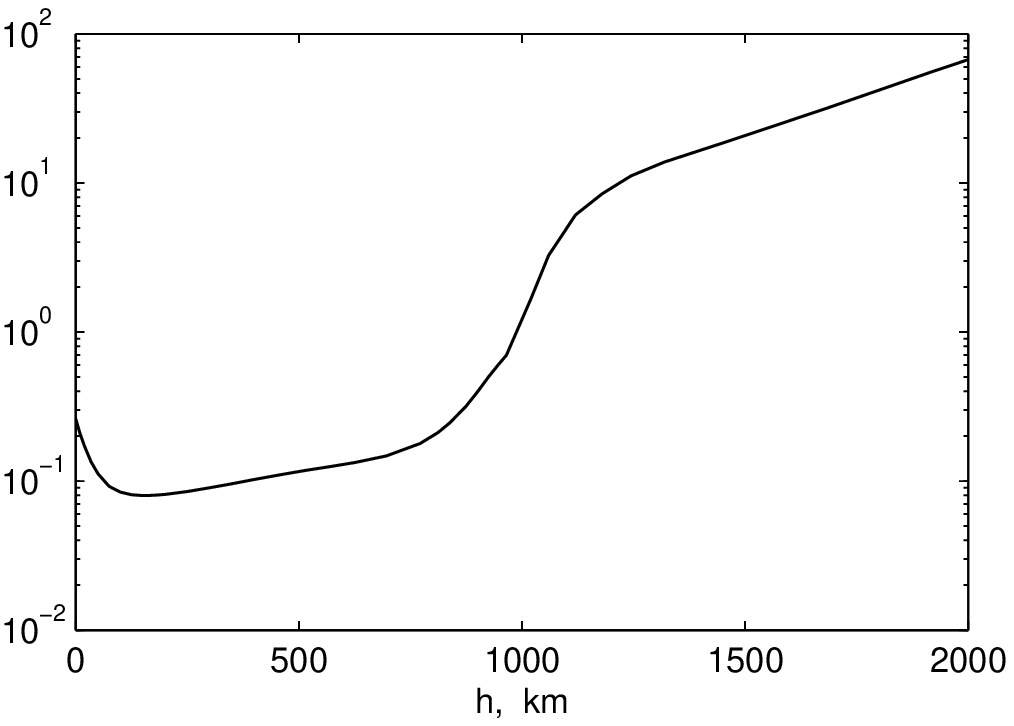}
\caption{The ratio of Coulomb and electron-neutral collision rates $N$ as a function of height.}
\label{fig:1}
\end{figure}

The terms proportional to $N=\nu _{ep}/\nu _{en}$ in Eqs.~(\ref{eq:5}-\ref{eq:6})
describe the effect of the electron-ion (Coulomb) collisions on the
FBI and represent the main analytical result of this letter. When the Coulomb
collisions are neglected ($N=0$), then the Eqs.~(\ref{eq:5}-\ref{eq:6}) reduce to
the well known result of \citet{FPF84}, which indicates that
in plasmas with $\kappa >1$, the FBI cannot develop regardless of the neutral drag velocity. In contrast,
if the Coulomb collisions are sufficiently frequent, the
FBI can appear even when the ions are relatively highly magnetized. The
dependence of $N$ on height, based on data of a semi-empirical chromospheric
model SRPM 306 \citep{FBH07}, is shown in Fig.~\ref{fig:1}. It is seen that the Coulomb collisions dominate
the electron-neutral collisions ($N>1$) in the upper half of the solar chromosphere,
at heights $h>1000~\mathrm{km}$. At these heights they cannot be ignored and
facilitate the FBI.

In the absence of Coulomb collisions, the physical background of the FBI is as follows
 \citep{OMO96,SN}: the electric field perturbation, $\delta \mathbf{E}%
\parallel \mathbf{k}$ is the leading force acting on the ions and causes both
the Hall ($\delta \mathbf{V}_{i}^{\mathrm{Hall}}\sim \delta \mathbf{E}\times
\mathbf{B}_{0}$) and the Pedersen ($\delta \mathbf{V}_{i}^{\mathrm{Ped}}\sim
\delta \mathbf{E}$) ion drift velocities. The FBI is powered by
the force due to convective term $(\mathbf{V}_{0i}\cdot \nabla )(\delta
\mathbf{V}_{i}^{\mathrm{Ped}}+\delta \mathbf{V}_{i}^{\mathrm{Hall}})$ in the
ion momentum equation (\ref{eq:2}). This force, acting on the ions, causes the
Hall and Pedersen responses in the perturbed ion velocity. The total
velocity response parallel to $\mathbf{k}$ contains two parts: the Pedersen
response (due to the Pedersen velocity $\delta \mathbf{V}_{i}^{\mathrm{Ped}}$)
giving rise to the destabilizing term in (\ref{eq:6}), and the Hall response
(due to the Hall velocity $\delta \mathbf{V}_{i}^{\mathrm{Hall}}$) giving rise
to the stabilizing term proportional to $\kappa ^{2}$.

Our analysis shows that, in the case of relatively frequent Coulomb
collisions, the physics of the FBI is significantly modified. The reason for this
is that the Hall and the Pedersen responses are both influenced by the Coulomb collisions
but in a different way: the Coulomb collisions reduce the Pedersen response but not so
much as the Hall response. This circumstance favors the FBI and makes it possible even for
$\kappa >1$.

\begin{figure}
\epsscale{.80}
\plotone{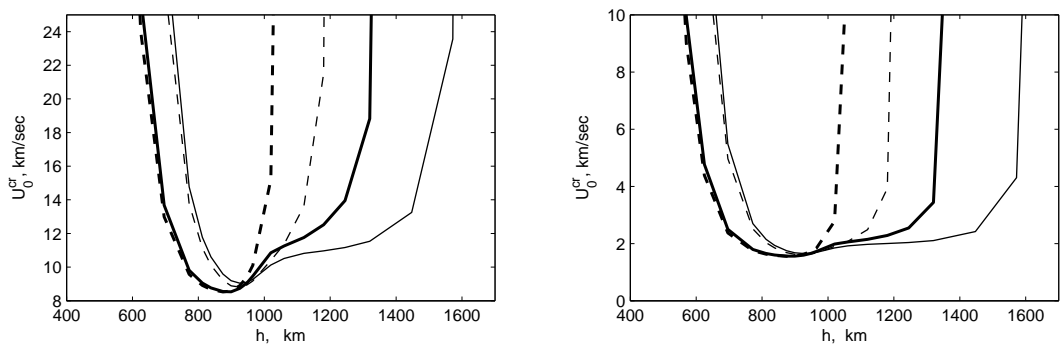}
\caption{Dependence of the threshold value of the velocity $U_0^{cr}$ on height with (solid lines) and without (dashed  lines) Coulomb collisions, for $B=30 {\rm G}$ (thin lines) and for $B=60 {\rm G}$ (thick lines). Left panel corresponds to the protons and right panel to ions with $m_i=30m_p$.} \label{fig:2}
\end{figure}

We can determine the threshold value of the relative velocity $U_{0}^{\mathrm{cr}}$ necessary to trigger the FBI in the framework of the model SRPM 306. Fig.~\ref{fig:2} shows the dependence of $U_{0}^{cr}$ on height with (solid lines) and
without (dashed lines) Coulomb collisions, for $B_{0}=30$ $\mathrm{G}$ (thin
lines) and for $B_{0}=60$ $\mathrm{G}$ (thick lines). The left panel corresponds
to the protons and the right panel to the ions with $m_{i}=30m_{p}$.

The results obtained here allow us to draw conclusions about the possible role of the
FBI in the inter-network chromospheric heating. The threshold value of the
current velocity necessary to trigger the FBI corresponds to the current density $%
J_{0}=en_{0e}U_{0}^{cr}$. Even for the lower chromosphere, where
the positively charged particles are mainly heavy ions and $U_{0}^{cr}\sim 2%
\mathrm{km/sec}$, very strong current densities $J_{0}\sim
2.4\times 10^{6}~\mathrm{statampere/cm^{2}}$ are required for the FBI to
develop. According to recent observations of \citet{S07}, at length-scales
of the order $100~\mathrm{km}$ and higher, the typical values of the
observed currents are much smaller, $\sim 5\times 10^{4}~\mathrm{%
statampere/cm^{2}}$. It is in principle possible that such strong currents
could exist at smaller scales. However, as we show below, in this case the
heating rate produced by the frictional dissipation of the relative
ion-neutral motion would be much higher than the power required to sustain
the radiative loses in the chromosphere. Indeed, the rate of frictional
dissipation in partially ionized plasmas is \citep{B65}
\begin{equation}
Q_{fr}=m_{e}n_{e}\nu _{ei}(\mathbf{V}_{e}-\mathbf{V}_{i})^{2}+m_{e}n_{e}\nu
_{en}(\mathbf{V}_{e}-\mathbf{V}_{i})^{2}+m_{p}n_{e}\nu _{ei}(\mathbf{V}_{e}-%
\mathbf{V}_{i})^{2},  \label{eq:7}
\end{equation}%
where the terms on the right-hand side are due to electron-ion,
electron-neutral and ion-neutral frictions, respectively. By substituting here
Eqs.~(\ref{eq:3a})-(\ref{eq:3}), we obtain
\begin{equation}
Q_{fr}=m_{e}n_{e}\nu _{en}U_{0}^{2}\left( 1+\kappa ^{2}+N+\frac{1}{\psi }%
\right) .  \label{eq:7a}
\end{equation}%
At relatively low heights in the solar chromosphere, where the positive charge is
dominated by heavy ions, the friction dissipation is dominated by ion-neutral
friction (the term proportional to $1/\psi $). Even for heights around $%
h=850~\mathrm{km}$, where $U_{0}^{cr}\sim 2~\mathrm{km/sec}$ and the
electron density is relatively small in accordance to the SRPM 306, we have $%
Q_{fr}\sim 40~\mathrm{erg/cm^{3}~sec}$. In terms of the associated
energy flux, such a  heating rate would be produced by a wave
flux $F\sim Q_{fr}H\sim 4\times 10^{8}~\mathrm{erg/cm^{2}~sec}$
dissipated in the lower chromosphere with $H\sim 100~\mathrm{km}$
being the characteristic width of the unstable layer. This value
of the flux is at least one order of magnitude higher than
necessary to compensate the radiative loses of the chromosphere
and such fluxes are not observed in the chromosphere. It is not
impossible, that the strong over-threshold currents occur
sporadically at small scales, and drive sporadic FBI events at the
length scales unresolvable for modern observations. However, given
the strong frictional heating associated with such currents, the
role of the FBI in the current dissipation and the associated
heating is of minor importance. On the other hand, the small-scale
plasma irregularities produced by the FBI can scatter radio waves,
and hence provide a diagnostic tool for strong cross-field
chromospheric currents if they exist. This last issue requires
further investigation.

In particular, our preliminary analysis of the full dispersion
equation (\ref{eq:4}) has already demonstrated that for typical
chromospheric parameters the strongest FB instability occurs at a
characteristic wavelength that varies with height in the range 0.1-3
m. In the middle chromosphere at 1000 km height the maximum instability growth rate $%
\gamma \approx 2\times 10^{3}$ s$^{-1}$ occurs for waves
with characteristic wavelengths $\lambda \approx 16$ cm. The
decimetric radio emission should effectively interact with the
electron density irregularities produced by the FB instability,
and this should result in observable scintillations of the decimetric
radio emission.

In summary, we determined the threshold current velocity for the FBI to occur
in a weakly ionized plasma taking into account
the finite ion magnetization and the electron-ion Coulomb collisions. We have
shown that, in the presence of Coulomb collisions, the FBI can occur
even when the ion magnetization is greater than unity. Applying
these analytical results to the solar chromosphere, we concluded that the FBI cannot be
responsible for the chromospheric heating at global length scales. The FBI
at small length scales cannot be excluded,
but the heating produced by the FBI cannot compete with the frictional heating under
chromospheric conditions. The small-scale irregularities generated by the FBI
can be used for remote
diagnostics of strong cross-field currents in the solar chromosphere.

\acknowledgments
Authors are grateful to Juan Fontenla for providing SRPM 306 data
and to Jovo Vranjes for helpful discussions.

G.G.\ acknowledges the hospitality of the Abdus Salam ICTP where part of the work was done
and partial support from INTAS grant 061000017-9258
and Georgian NSF grants ST06/4-096 and ST07/4-193.
These results were obtained in the framework of the projects
GOA/2009-009 (K.U.Leuven), G.0304.07 (FWO-Vlaanderen) and
C~90205 (ESA Prodex 9). Financial support by the European Commission through the SOLAIRE
Network (MTRN-CT-2006-035484) is gratefully acknowledged.





\clearpage



\end{document}